\documentclass[conference]{IEEEtran}
\IEEEoverridecommandlockouts
\usepackage{cite}
\usepackage{amsmath,amssymb,amsfonts}
\usepackage{algorithmic}
\usepackage{graphicx}
\usepackage{textcomp}
\usepackage{xcolor}
\usepackage{booktabs}
\usepackage{multirow}
\usepackage{subfigure}
\usepackage{url}
\usepackage{stfloats}
\usepackage{float}
\IEEEoverridecommandlockouts

\def\BibTeX{{\rm B\kern-.05em{\sc i\kern-.025em b}\kern-.08em
    T\kern-.1667em\lower.7ex\hbox{E}\kern-.125emX}}
\begin{document}

\title{MRC-LSTM: A Hybrid Approach of Multi-scale Residual CNN and LSTM to Predict Bitcoin Price\\
\thanks{$\ast$ indicates the corresponding author.}
}

\author{\IEEEauthorblockN{Qiutong Guo}
\IEEEauthorblockA{\textit{School of Computer Science} \\
\textit{Sichuan University}\\
Chengdu, China}
\and
\IEEEauthorblockN{Shun Lei}
\IEEEauthorblockA{\textit{School of Computer Science} \\
\textit{Sichuan University}\\
Chengdu, China }
\and
\IEEEauthorblockN{Qing Ye$^{\ast}$}
\IEEEauthorblockA{\textit{School of Computer Science} \\
\textit{Sichuan University}\\
Chengdu, China }
\and
\IEEEauthorblockN{Zhiyang Fang}
\IEEEauthorblockA{\textit{School of Computer Science} \\
\textit{Sichuan University}\\
Chengdu, China }
}
 \maketitle

\begin{abstract}
Bitcoin, one of the major cryptocurrencies, presents great opportunities and challenges with its tremendous potential returns accompanying high risks. The high volatility of Bitcoin and the complex factors affecting them make the study of effective price forecasting methods of great practical importance to financial investors and researchers worldwide. In this paper, we propose a novel approach called MRC-LSTM, which combines a Multi-scale Residual Convolutional neural network (MRC) and a Long Short-Term Memory (LSTM) to implement Bitcoin closing price prediction. Specifically, the Multi-scale residual module is based on one-dimensional convolution, which is not only capable of adaptive detecting features of different time scales in multivariate time series, but also enables the fusion of these features. LSTM has the ability to learn long-term dependencies in series, which is widely used in financial time series forecasting. By mixing these two methods, the model is able to obtain highly expressive features and efficiently learn trends and interactions of multivariate time series. In the study, the impact of external factors such as macroeconomic variables and investor attention on the Bitcoin price is considered in addition to the trading information of the Bitcoin market. We performed experiments to predict the daily closing price of Bitcoin (USD), and the experimental results show that MRC-LSTM significantly outperforms a variety of other network structures. Furthermore, we conduct additional experiments on two other cryptocurrencies, Ethereum and Litecoin, to further confirm the effectiveness of the MRC-LSTM in short-term forecasting for multivariate time series of cryptocurrencies.
\end{abstract}

\section{Introduction}
Bitcoin is the world's first distributed super sovereign digital currency, proposed and established by Satoshi Nakamoto \cite{nakamoto2008peer} in 2009. It relies on an electronic payment system based on cryptography and P2P (Point to Point) technology. By using encryption algorithms and automatic authentication mechanisms, it is difficult to crack or forge, making its transactions more secure and transparent. Since the birth of Bitcoin, it has quickly gained widespread attention.

As a new type of cryptocurrency, Bitcoin conducts 24/7 transactions and is able to exchange for many major currencies at a low cost of foreign exchange. Compared to other traditional financial assets, Bitcoin provides investors a new type of portfolio management. Existing research \cite{Dyhrberg2016,DYHRBERG2016139} indicated that Bitcoin has an apparent role in the portfolio management market. Empirical analysis by Anne H et al. \cite{DYHRBERG2018140} also corroborates the investability of Bitcoin. In recent years, its rich portfolio and the potential for high returns have attracted the attention of an increasing number of financial investors. However, the Bitcoin market is highly volatile and subject to frequent speculative bubbles \cite{bubbles2015,CORBET201881}, thus its trading risks are enormous. Consequently, as discussed above, the study of effective price forecasting methods is of great practical importance to investors, researchers, and policymakers around the world due to the tremendous opportunities and challenges posed by Bitcoin.

In recent years increasingly machine learning methods, especially deep neural networks(DNNs), have been applied to market forecasting in cryptocurrencies. In terms of the Bitcoin market, it is highly volatile and generates a large amount of highly non-linear transaction data. In order to effectively explore these dynamic data, we need models which are able to analyze the internal interactions and hidden patterns in the data. Several researchers \cite{ADCOCK2019121727,NAKANO2018587} have shown that DNN models are well suited for learning lagged correlations between step-wise trends in large financial time series. In a literature review \cite{KHASHEI2010479} of comparative studies between artificial neural networks and traditional statistical models, it was found that in 72\% of 96 cases, artificial neural network models were shown to have better predictive performance. Consequently, there is no doubt that due to the highly nonlinear and volatile nature of financial markets, DNN models are increasingly applied in the field of finance, especially for financial time series forecasting.
 
In this paper, a novel DNN model consisting of a Multi-scale Residual Convolutional Neural Network with Long Short-Term Memory (LSTM) is proposed for Bitcoin price prediction. In the study, not only the influence of transaction information such as Bitcoin historical price is considered on the closing price of Bitcoin, but also external influences such as macroeconomic factors and investors’ attentions are introduced. The proposed model consists of two main parts. First, is the proposed multi-scale residual module based on one-dimensional convolution. It contains an identity mapping and three branching networks, where the size of the convolutional kernel is different for each branching network. Then, the second part is the LSTM network, which is able to further learn the trends of multivariate time series and the interactions between the series, and output the final predicted values. Different from previous financial multivariate time series forecasting, we construct a multi-scale residual module, which can also be called a three-bypass residual module, in which information from these bypasses can be shared with each other. Moreover, this structure can extract potential features that have a high impact on bitcoin price in multiple time scales and integrate them into highly expressive feature vectors after the concatenate operation. In addition to the Bitcoin price prediction experiments, we also conducted extensive experiments on datasets of two different cryptocurrencies (Litecoin and Ethereum) to confirm the strong ability of the proposed MRC-LSTM model for short-term price prediction of cryptocurrencies.
 
The remainder of the paper is organized as follows. Section \ref{sec:related-work} reviews the relevant literature. Section \ref{sec:MRC-LSTM} introduces the methodology, including the proposed model. Section \ref{sec:experiments} presents our experiments, while our empirical results and discussion are shown in Section \ref{sec:dis}. Finally, Section \ref{sec:conclusion} is the conclusion of this paper.

\section{Related literature}
\label{sec:related-work}
In this section, the related work about price prediction of the cryptocurrency and the basis knowledge to the MRC-LSTM are provided.
\subsection{Neural Network Approaches}
Salim Lahmiri et al.~\cite{lahmiri2019cryptocurrency} first applied the deep neural networks (DNNs) to cryptocurrency price prediction, elucidated the short-term predictability of cryptocurrencies, and found that the predictive accuracy of Long Short-term Memory (LSTM) neural networks is higher than that of Generalized Regression Neural Networks (GRNN)\cite{specht1991general}. Subsequently, a growing body of research has emerged in this area \cite{Liang_2020,zhu2021retrieving}. LSTM neural networks \cite{hochreiter1997long} have been shown to be significantly effective in forecasting bitcoin prices due to their ability to identify long-term dependencies and store both long-term and short-term temporal information. Researchers have obtained better results with the LSTM network, in their works to predict bitcoin prices and compare them with the performance of different models \cite{uras2020forecasting,linardatos2020bitcoin}.

Further, Convolutional Neural Networks (CNNs) have also been applied to cryptocurrency market forecasting. In a study \cite{ALONSOMONSALVE2020113250}, researchers combined CNN with LSTM neural network for high-frequency market trend prediction for a variety of cryptocurrencies. Their empirical analysis shows that the addition of convolutional layers improves the prediction performance and that the hybrid network structure provides the best prediction results in the experiment. Yan Li et al. \cite{CNNLSTM2020} propose a hybrid neural network model based on CNN and LSTM neural networks, and experimental results show that CNN-LSTM hybrid neural network can effectively improve the accuracy of value prediction and direction prediction compared to single-structure neural network. The hybrid neural network model combining CNN with LSTM has also been used by many researchers for time series prediction of financial data such as Gold volatility prediction\cite{VIDAL2020113481}, stock prices, etc. Its excellent performance has also been demonstrated in many other fields \cite{tian2018deep,ijcai2017-562,lei-etal-2018-sequicity}.

\subsection{Time Series Forecasting}
Time series research has always been an important field of machine learning. By building up neural networks based on deep learning, we are able to extract and exploit the hidden information represented by digital currency raw data of digital currencies in order to make accurate and efficient price predictions \cite{lahmiri2019cryptocurrency}. In recent years, researchers have continuously made progress in the field of financial time series forecasting \cite{ariyo2014stock,adhikari2014combination,cao2019financial}. Many traditional research approaches focus on learning internal patterns, such as autocorrelation, in a particular time series. However, for reality scenarios, especially in cryptocurrencies, stock markets and other financial domains, in most cases we are dealing with multivariate time series. Changwei Hu et al. \cite{hu2019deep} developed a deep learning structural time series model to handle correlated multivariate time series inputs. Their model is able to leverage dependencies among multiple correlated time series and extract weighted differencing features for better trend learning. The existence of close correlations among many multivariate time series motivates us to consider not only intra-series pattern learning but also inter-series pattern learning when dealing with such tasks.

\subsection{Residual Network}
As one of the milestones in the evolution of CNN, Residual Network (ResNet) \cite{he2016deep} has achieved impressive, record-breaking performance on many challenging tasks. ResNet share some similarities with Highway networks, such as residual blocks and shortcut connections. ResNet simplifies the training of very deep networks by bypassing signals from one layer to the next through identity connections. The basic idea underlying residual learning is the branching of gradient propagation paths. For CNNs, this idea was first introduced in the form of parallel paths in the inception models of \cite{szegedy2015going}. Many research works  \cite{long2015fully,yang2015multi} have exploited the multi-level features in CNNs by skip-connections and found them to be effective for a variety of visual tasks. GoogLeNet \cite{szegedy2015going,szegedy2016rethinking} proposed an “Inception module” that connects feature mappings generated by filters of different sizes to increase the diversity of feature extraction. Meanwhile, GoogLeNet increased the network width by skipping connections to improve the robustness and expressiveness of the network. Gao Huang et al. \cite{huang2017densely} proposed the Dense Convolutional Network (DenseNet) to exploit the potential of the network through feature reuse. In contrast to ResNet, it introduces a direct connection from any layer to all subsequent layers. In addition, DenseNet combines features by concatenating them, rather than combining features through summation. These works show us the utility of skip connections, and DenseNet also shows us how to connect feature maps via concatenation. Our work is partly inspired by these two ideas and explores their application to model building for short-term price forecasting of cryptocurrencies.

\section{Methodologies}
\label{sec:MRC-LSTM}
In this section, the details of our proposed method MRC-LSTM and basic blocks are described.
\subsection{ResNet}
\begin{figure}[htbp]
\centerline{\includegraphics[width=0.3\textwidth]{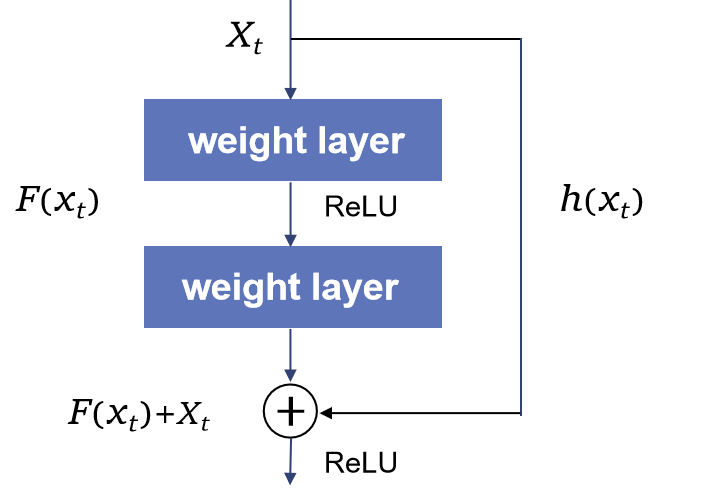}}
\caption{the residual block}
\label{fig:res}
\end{figure}

In order to solve the problem of gradient disappearance and gradient explosion due to network deepening, He et al. proposed a new DNN——ResNet\cite{he2016deep}. ResNet consists of many residual units, shown in Fig.~\ref{fig:res}, and each residual unit can be represented in the following Equations~\ref{res_1} and ~\ref{res_2}.

\begin{equation}\label{res_1}
    y_t =h(x_t) + F(x_t,w_t)
\end{equation}
\begin{equation}\label{res_2}
    x_{t+1} = f(y_i)  
\end{equation}
Where $F$ is a residual function, $f$ is a ReLU function, $w_t$ is the weight matrix, and $x_t$ and $y_t$ are the inputs and outputs of the $t$-th layer. The function $h$ is an identity mapping given by Equation~\ref{res_3}:
\begin{equation}\label{res_3}
    h(x_t)=x_t
\end{equation}

In a residual block, skip connections can effectively aggregate historical information, reduce the loss of features and information to some extent, and enable the network to learn richer content. Motivated by the idea of Residual units, our proposed model applies the method of skip connections.

\subsection{LSTM}
\begin{figure}[htbp]
\centerline{\includegraphics[width=0.45\textwidth]{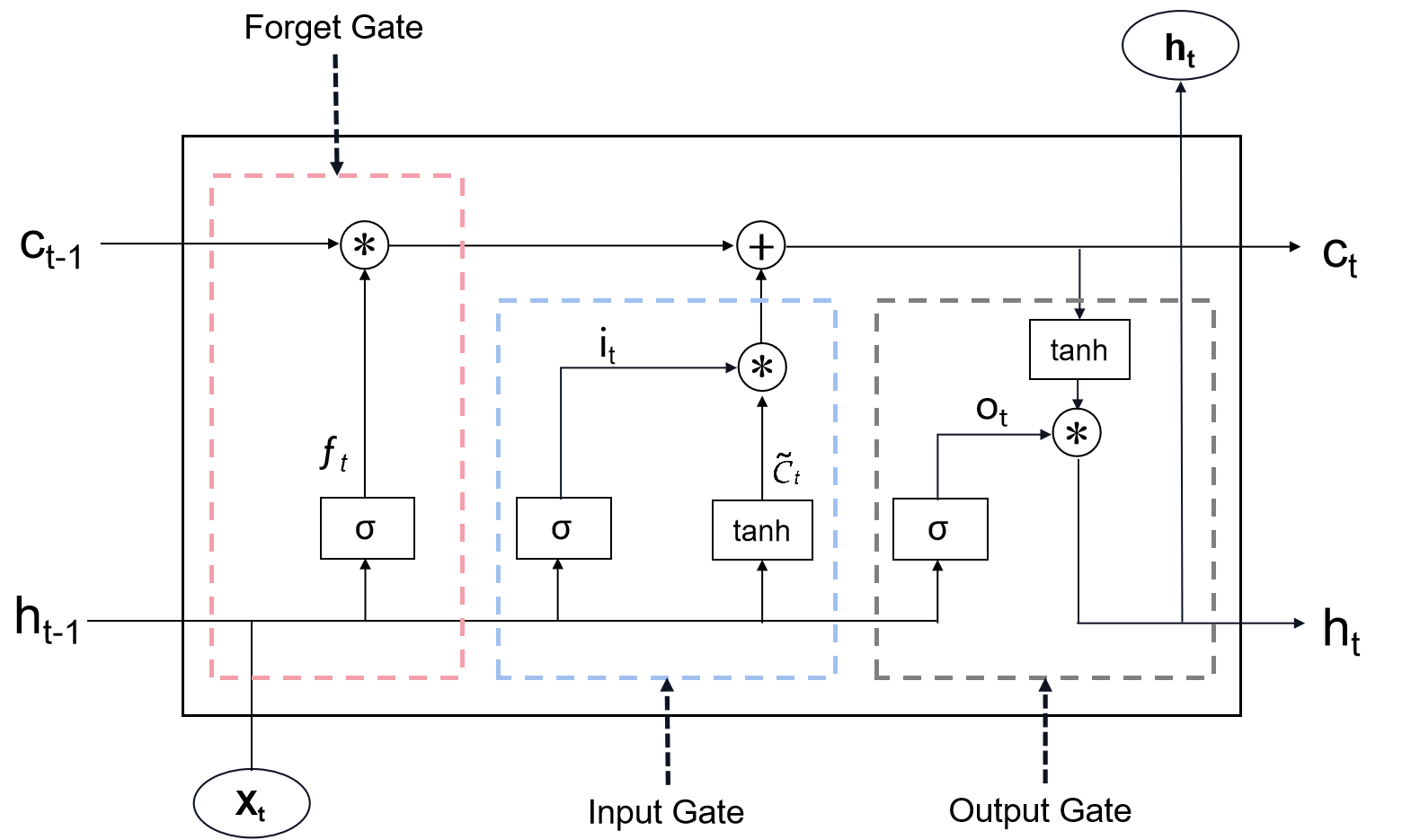}}
\caption{Basic unit of LSTM network}
\label{fig:lstm}
\end{figure}

LSTM is based on the Recurrent Neural Network (RNN) model, which can effectively solve the gradient disappearance and gradient explosion problems in the RNN model. The addition of a special gate control mechanism makes it possible to solve the problem of long-term dependence, and it is suitable for time series processing and prediction, natural language generation \cite{lei2018sequicity, jin2018explicit, pan2019recent}. Fig.~\ref{fig:lstm} gives the basic unit of the LSTM neural network. Its basic unit is the memory block, which contains the memory cell and three gates that control the memory cell, namely, the input gate, the output gate, and the forget gate.

Forget Gate is mainly used to calculate the degree of information retention and discarding. The output $f_t$ represents the probability of forgetting the state of the underlying cell layer, and is calculated as follows:
\begin{equation}
    f_t = \sigma(W_f x_t+U_f h_{t-1} + b_f)
\label{lstm1}
\end{equation}

In the Equation \eqref{lstm1}, the $x_t$ indicates the input of the current cell, $h_{t-1}$ presents the output of the previous cell, $\sigma$ is the sigmoid function.

In the input gate, the sigmoid activation function is used to calculate which information needs to be updated. Then, use the tanh activation function to get a vector of candidate values $\tilde{C}_t$, after which updates the previous state $C_{t-1}$ to $C_t$. The formulas are Equations~\ref{ingate1}, ~\ref{ingate2} and~\ref{ingate3}.

\begin{equation}
    i_t = (W_i x_t+U_i h_{t-1}+b_i)
\label{ingate1}
\end{equation}
\begin{equation}
    \tilde{C}_t = tanh(W_c x_t+U_c h_{t-1}+b_c)
\label{ingate2}
\end{equation}
\begin{equation}
    C_t = f_t\cdot C_{t-1}+i_t\cdot\tilde{C}_t
\label{ingate3}
\end{equation}

The output gate is used to calculate the extent of the information output at the current moment. The information is filtered by the sigmoid activation function to obtain $o_t$. Then the tanh activation function is used to obtain the desired information $h_t$:
\begin{equation}
   o_t = \sigma(W_o x_t+U_o h_{t-1}+b_o)
\end{equation}
\begin{equation}
    h_t = o_t\cdot tanh(C_t)
\end{equation}

\subsection{The Proposed Network Architecture.}
Inspired by ResNet, and based on the need to deal with time series problems, we developed a novel building block, Multi-scale Residual Convolutional block (MRC) based on one-dimensional time convolution. It is combined with an LSTM neural network to form a new hybrid network architecture (MRC-LSTM) to perform cryptocurrency time series prediction. 
In the following, we will first introduce how to design the multi-scale residual block, followed by the proposed MRC-LSTM model, i.e., how the new building block can be combined with LSTM to predict the bitcoin price.

 \begin{figure}[htbp]
\centerline{\includegraphics[width=0.4\textwidth]{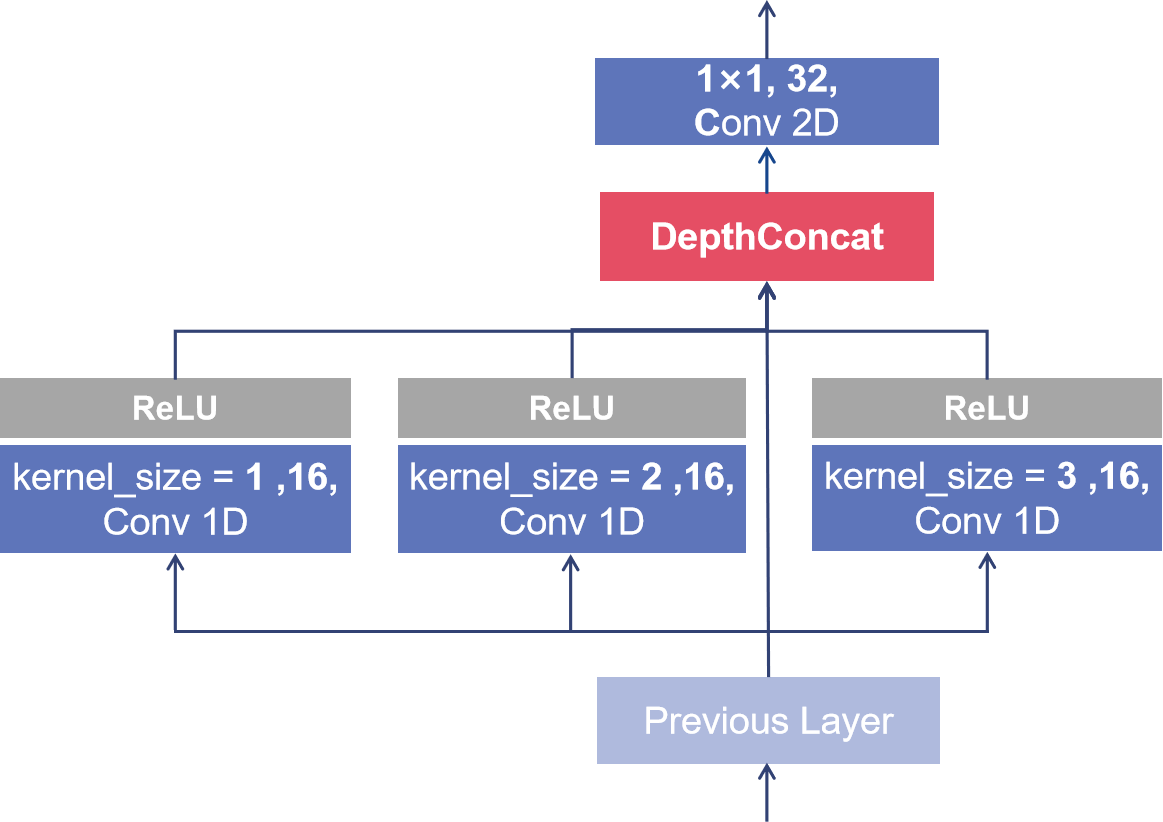}}
\caption{The Structure of Multi-scale Residual Block}
\label{fig:res_module}
\end{figure}

\textbf{The Multi-scale Residual Block.} This is the first part of MRC-LSTM which is utilized to extract potential features with high expressiveness at different time scales from the dataset. We constructed a three-bypass convolutional layer with convolutional kernels of different sizes, and joined a jump connection to further aggregate historical information efficiently. Due to the high volatility of Bitcoin and the short-term predictability of cryptocurrencies \cite{lahmiri2019cryptocurrency}, the Bitcoin market is suitable for short-term forecasting, and the longer the forecast period, the worse the forecast performance. Therefore, we select 1D convolutional kernels of size 1, 2, and 3 to slide the sequences in the temporal domain, which means that such multi-scale residual module can simultaneously extract the trends and the hidden interactions of the data within 1, 2, and 3 adjacent days in the sequences. In addition, it is known that the window size of the kernel in a one-dimensional convolution will affect the network learning effect. It is likely to miss local feature information when the window size is too large; When the window size is small, local features are easy to extract, but may reduce the correlation of local features. Hence, the "Multi-scale" design of the proposed model can also take into account the advantages of both large and small windows. Fig.~\ref{fig:res_module} shows the design of the residual module.

When it comes to the way to combine the extracted features of three bypasses, here we draw on the idea of DenseNet, which differs from ResNet in that we never combine features by summing them before they are passed to the next layer; instead, we combine features by concatenating them. We know that the concatenation operation is not viable when the feature-map size changes. Therefore, in order to keep the feature-maps size uniform, we need to perform a zero-padding when performing 1D convolution. Then, the identity mapping is concatenated in the depth direction with the feature maps of the three paths and fed into the subsequent layer.

At the end of the module, the use of 1 × 1 2D convolutional kernels enables cross-channel information interaction and expansion. In this way, the feature-maps extracted by the three prediction cycles will be fused and the network will adaptive extract useful information from these hierarchical features. Meanwhile, non-linearity can be added while keeping the feature map scale constant.

\begin{figure*}[htbp]
\centerline{\includegraphics[width=0.95\textwidth]{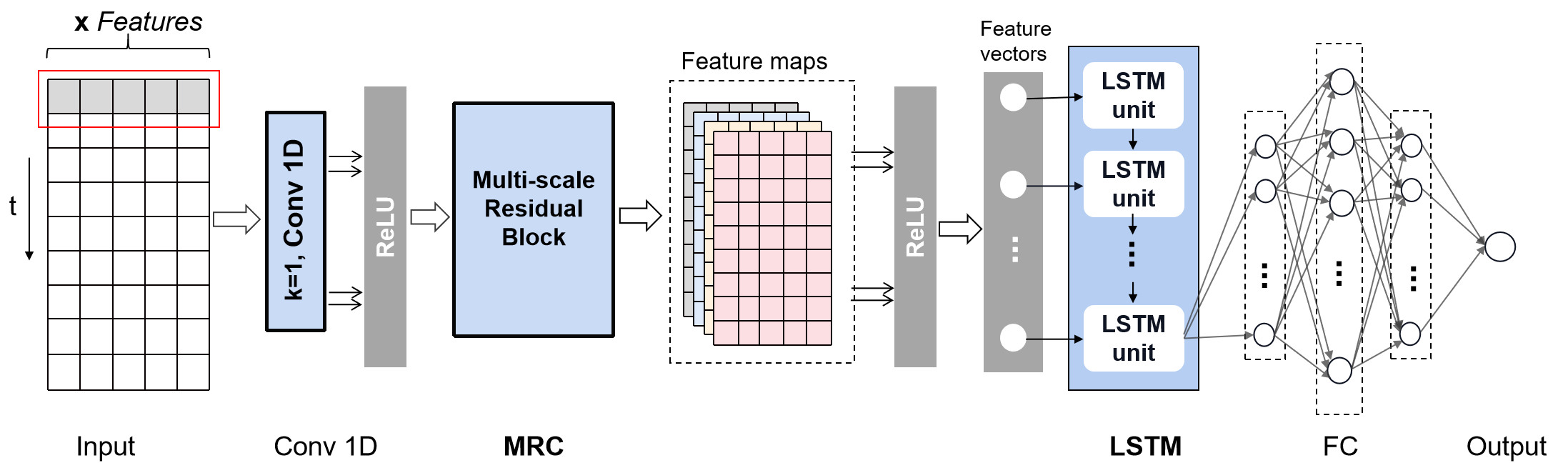}}
\caption{The overall architecture of MRC-LSTM model. Conv 1D, One-dimensional convolutional layer; MRC, multi-scale residual block; LSTM, long short-term memory; FC, fully connected layer.}
\label{fig:proposed_arch}
\end{figure*}

\textbf{Hybrid MRC-LSTM model.} The network consists of two main parts, the first is the multi-scale residual module for extracting features in the multivariate time series, and the second is the LSTM layer for learning pattern changes and predicting prices. Fig.~\ref{fig:proposed_arch} shows the structure of the MRC-LSTM neural network. The network contains an input layer, a 1D convolutional layer, the multi-scale residual module, an LSTM layer, a fully connected layer, and an output layer. 

During the construction of the MRC-LSTM, we first use a set of 1D convolutional kernels with a window size of 1 to slide in the time direction for initial feature extraction and feature augmentation of the sequence. Then, the multi-scale residual module is utilized to extract and integrate local features across multiple time scales. Finally, an LSTM network is used to learn relationships in time steps for price forecasting.
 \begin{equation}\label{relu}
     f(x) = \begin{cases}
x , (x>0)\\
0 , (x\le0)
\end{cases}   
\end{equation}

We introduce activation functions to incorporate nonlinearities in the network. The common activation functions include Sigmoid, Tanh, ReLU, and SELU. In this paper, the rectified linear unit (ReLU) is selected as the nonlinear activation function, which is expressed as the Equation~\eqref{relu}. 

\section{Experiments}
\label{sec:experiments}
\subsection{Data Collection and pre-processing}
The dataset used in this experiment includes the daily closing price of Bitcoin and 10 types of internal (Bitcoin trading datas) and external (macroeconomic variables and investor attention) information that have an impact on the price of Bitcoin from October 25, 2015 to October 17, 2020, for a total of 1820 records. The dataset is divided into 80\% as training set and 20\% as test set. The training set is further divided into a training set (80\%) and a validation set (20\%) for evaluating performance and avoiding overfitting. The Table~\ref{attribute} shows all the attributes that compose the dataset.

The internal information in the dataset refers to the daily transaction datas of Bitcoin, including the open price, the closing price, the highest price, the lowest price, the weighted price, the trade volume in Bitcoins and the market's currency. They are collected from the offical website \footnote{ https://bitcoincharts.com}.

As for external factors, we take the macroeconomic factors and investor attention into consideration. As a new type of investment instrument, the price of bitcoin is considered to be related to several macroeconomic variables. The study by Leon Li considered four representative volatility indices published by the Chicago Board Options Exchange(CBOE), including the Volatility Index(VIX) and the Gold Price Volatility Index(GVZ). He believed that volatility indices can be used to speculate and trade on market sentiment regarding future volatility \cite{li2020risk}. Some researchers have found that there is a relationship between the price of bitcoin and gold, crude oil, and stock market indices \cite{vassiliadis2017bitcoin}. Therefore, this paper introduces macroeconomic variables that have been proven to have predictive power for the bitcoin market to predict bitcoin prices. Ultimately, the following macroeconomic factors were chosen for the experiment: S\&P 500 Index, GVZ, VIX. They are all daily data from the Wind Database.

Moreover, in recent years, a number of researchers have found that Google trends, which reflect investors' attention, play a significant role in predicting the Bitcoin market \cite{urquhart2018causes,yelowitz2015characteristics}. Da et al. \cite{da2011search} pointed out that search activity can reflect investors' attention. Therefore, we employ the Google Trends as one of the predictors, which is derived from Google Trends~\footnote{ https://trends.google.com}.

\begin{table}[htbp]
\scriptsize
\centering
\caption{List of input attributes}
\label{features}
\begin{tabular}{ccc}
\toprule
\textbf{Transaction Information}
& \textbf{Macroeconomic Variables}  &  \textbf{Investor Attention}\\
\midrule
the Open price &  S\&P 500 Index & Google Trends\\
the Close price & GVZ & -\\
the Highest price & VIX & -\\
the Lowest price & - & - \\
the Weighted price & - & - \\
Volume(BTC) & - & - \\
Volume(Currency)&- & - \\
\bottomrule
\end{tabular}
\label{attribute}
\end{table}

The prediction period used in this experiment is five days, which means the closing price of bitcoin on the sixth day is predicted using the characteristic parameter data from the previous five days. In the meantime, our proposed residual module is capable of using three convolutional kernels of different scales to extract the temporal characteristics of a multivariate sequence between one, two, and three adjacent days in a 5-day period. The richer potential information contained in the sequences is extracted by getting different sized receptive fields.

The original data is usually normalized to eliminate scale effects between indicators before modeling. The common normalization methods are min-max normalization and Z-Score. In the experiment, the min-max normalization is used. This approach, as shown in Equation~\eqref{min-max}, is capable of varying the original data linearly and mapping the values of the data between 0 and 1.
\begin{equation}
   x_{norm} = \frac{x_t-x_{min}}{x_{max}-x_{min}}
\label{min-max}
\end{equation}
where $x_t$, $x_{norm}$ are the input sample data and the data after normalization, $x_{min}$,$x_{max}$ are the minimum and maximum value of the samples. After modeling, the target outputs need to be anti-normalized.
\begin{equation}
   \hat{y}_t = y_{norm}(y_{max}-y_{min})+y_{min}
\end{equation}
where $y_t$ is the output of the prediction value, $y_{min}$, $y{max}$ are the minimum and maximum value of the target data, and $y_{norm}$ is the predicted value derived directly from the building model.

\begin{figure*}
    \centering
    \subfigure[MLP, MRC-LSTM and Actual curves]{
        \includegraphics[width=0.31\textwidth]{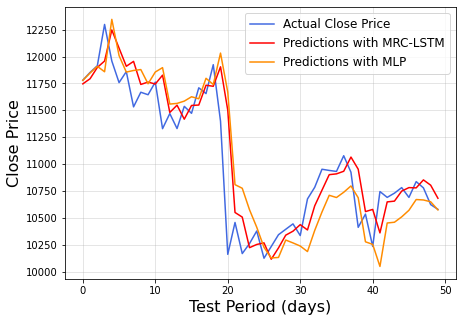}
    }
    \subfigure[LSTM, MRC-LSTM and Actual curves]{
        \includegraphics[width=0.31\textwidth]{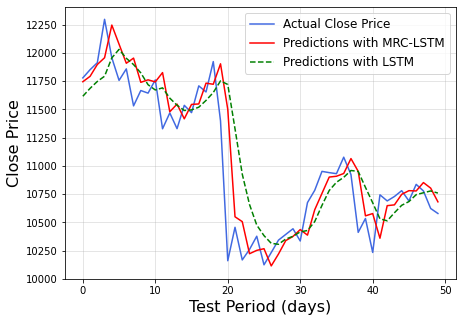}
    }
    \subfigure[CNN-LSTM, MRC-LSTM and Actual curves ]{
        \includegraphics[width=0.31\textwidth]{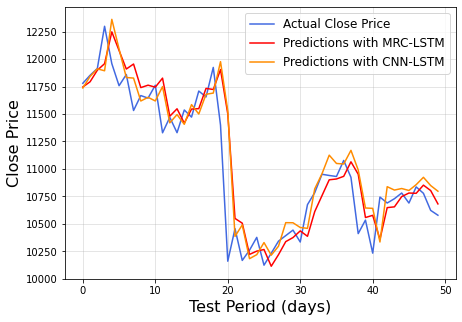}
    }
    \caption{Comparison diagrams of prediction and actual curves}
    \label{fig:results curves}
\end{figure*}

\subsection{Process and Parameters Settings}
The proposed neural network structure in this paper consists of two main parts. Through repeated experiments, the parameters of the proposed network are determined as follows.

\begin{itemize}
    \item In the first part, a 1D convolutional layer is used for initial feature extraction and expansion. There are 16 convolutional kernels of size 1 in this layer. After that, the features obtained from the upper layer are input into the proposed residual module for feature extraction and integration at multiple time scales. The parameters of the convolutional layer in the residuals module are shown in Fig.~\ref{fig:res_module}.
    
    \item In the second part of the network structure, the features extracted by the residual module are input into the LSTM layer, which has a layer number of 1 and a number of neurons of 50. Since the features extracted by the 1D convolution-based residual module are along the temporal dimension, they can be entered into the LSTM to learn the long-range time pattern changes of the trends in the sequence. Finally, the full connection layer is used and the predicted value is output.
\end{itemize}

In the experiment, the loss function is the MSE, the optimization function is the Adam optimization algorithm, the batch normalization number Batch size is 50. In addition, we use the Piece wise decay method for learning rate decay, with an initial learning rate of 0.001, and the learning rate decreases to 0.3 times the original rate every 500 epochs, for a total of 2000 epochs. Each time the neural network outputs a value, a sliding window is used for prediction. The experiments are compared and evaluated with actual values to obtain the final predictive performance. All algorithms are implemented using Pytorch~\footnote{https://pytorch.org/}.

\subsection{Performance Evaluation Criteria}
In this work, four error functions, Mean Absolute Error (MAE)(i.e., Equation~\eqref{MAE}), Root Mean Square Error (RMSE)(i.e., Equation~\eqref{RMAE}), Mean Absolute Percentage Error (MAPE)(i.e., Equation~\eqref{MAPE}) and the coefficient of determination($R^2$), are introduced as performance metrics to quantify the ability of the DNN to forecast prices.
\begin{equation}
\label{MAE}
   MAE = \frac{1}{N}\sum_{i=1}^N \left\vert(y_i-f_i)\right\vert 
\end{equation}
\begin{equation}
\label{RMAE}
   RMSE = \sqrt{ \frac{1}{N}\sum_{i=1}^N (y_i-f_i)^2}
\end{equation}
\begin{equation}
\label{MAPE}
   MAPE = \frac{100}{N}\sum_{i=1}^N \left\vert\frac{(y_i-f_i)}{y_i}\right\vert 
\end{equation}
where $y_i$ and $f_i$ are the $i$th  actual value and predicted value, and $N$ is the number of test samples. They reflect the deviation between the predicted and actual values. The larger their value, the greater the error in the forecast.

The $R^2$ value reflects the accuracy of the model which ranges from 0 to 1 with 1 denotes perfect match:
\begin{equation}
   R^2 =  \frac{\begin{matrix} \sum_{n=1}^n (\hat{y}_i-\Bar{y})^2  \end{matrix}}{\begin{matrix} \sum_{n=1}^n (y_i-\Bar{y})^2 \end{matrix}}
\end{equation}
where $\hat{y}_i$ represents the predicted value, $\Bar{y}$ is the average value, and $y_i$ is the observed value.

\begin{table}[htbp]
\scriptsize
\centering
\caption{Bitcoin: The errors of different deep neural networks}
\begin{tabular}{ccccc}
\toprule
{\textbf{Architecture}} 
&  \textbf{MAE}  &  \textbf{RMSE}   &   \textbf{MAPE}  &  \textbf{$R^2$(\%)}  \\
\midrule
MLP & 246.55 & 345.50 & 2.31 & 87.95 \\
LSTM & 212.71 & 317.24 & 1.99 & 89.83 \\
CNN & 191.92 & 268.26 & 1.77 & 92.73 \\
CNN-LSTM & 176.79 & 270.66 & 1.66 & 92.60 \\
MRC-LSTM & \textbf{166.52} &\textbf{ 261.44 }& \textbf{1.56} & \textbf{93.10} \\
\bottomrule
\end{tabular}
\label{error1}
\end{table}

\begin{table}[htbp]
\scriptsize
\centering
\caption{ETH and LTC: The errors of different deep neural networks}
\begin{tabular}{ccccccc}
\toprule
\multirow{2}{*}{\textbf{Architecture}} &
\multicolumn{3}{c}{\textbf{Ethereum(ETH)} } &
\multicolumn{3}{c}{\textbf{Litecoin(LTC)} } \\
\cmidrule(r){2-4} 
\cmidrule(r){5-7}
&\textbf{MAE}  &  \textbf{RMSE}   &   \textbf{MAPE} & 
\textbf{MAE}  &  \textbf{RMSE}   &   \textbf{MAPE} \\
\midrule
MLP & 1.04 & 1.53 & 7.83 & 0.42 & 0.70 & 9.67 \\
LSTM & 0.76 & 1.17 & 5.74 & 0.32 & 0.38 & 7.65 \\
CNN & 0.81 & 1.23 & 5.92 & 0.39 & 0.44 & 9.37 \\
CNN-LSTM & 0.77 & 1.24 & 6.20 & 0.27 & 0.37 & 6.43 \\
MRC-LSTM & \textbf{0.70} &\textbf{ 1.13 }& \textbf{5.43} & \textbf{0.14} & \textbf{0.26} & \textbf{3.17}\\
\bottomrule
\end{tabular}
\label{error2}
\end{table}

\section{Results and discussion}
\label{sec:dis}

The algorithms used for comparison in the experiments are Multilayer Perceptron (MLP), Two-dimensional Convolutional Neural Network (CNN), Long short-term memory (LSTM), and CNN-LSTM. Table~\ref{error1} shows the accuracy metrics of the proposed model compared to the benchmarks using different architectures.

From  Table~\ref{error1}, it is apparent that the MLP has the largest prediction error, while the LSTM outperforms the MLP. The LSTM can solve the long-term dependence problem in time series samples, making it more suitable for solving time series prediction problems than the MLP.

The CNN-LSTM model consists of a two-dimensional convolutional layer for feature extraction first, and then the extracted features are fed into the LSTM for price prediction. The prediction accuracy of the LSTM model is just lower than that of the CNN-LSTM, which illustrates that when there are many features and few samples, the LSTM cannot fully extract the important potential information from the data, which will affect the training effect of the neural network to some extent.

The error of MRC-LSTM prediction results is smaller than that of CNN-LSTM, which is the best prediction performance algorithm in this experiment. It indicates that the feature extraction strategy of Multi-scale residual block is more efficient. The main reason is that, on the one hand, in the one-dimensional convolutional layer, the convolutional kernel slides only along the time dimension and performs the convolution operation on all feature values at the time step, which reduces the problem of multiple covariance in the features. On the other hand, extracting features in multivariate sequences from three different scales improves the expression of local features and facilitates LSTM to make more accurate and fast predictions.

Besides, Fig.~\ref{fig:results curves} shows the effect of our proposed network by extending the prediction map to several sample points, and three sets of comparisons are made. It can be directly reflected from the figure that the predicted values derived from the MRC-LSTM model are overall closer to the actual values compared to several other models. The main reason is that the proposed residual module enables the model to combine the features extracted from the underlying convolution into the upper convolution and learn the trends embedded in the multivariate time series from different time scales. Consequently, it can make efficient use of the features in the series and improve the prediction performance of the model.
 
It is evident from the Fig.~\ref{fig:results curves} that the predictions of both the MRC-LSTM and several other models have some degree of lag with the actual values. This is probably because despite the outstanding performance of the Multi-scale residual convolution module in feature extraction as described above, we still cannot ignore the working principle of the LSTM, which uses information from previous lags to predict future instances. Moreover, as the Bitcoin market is a highly dynamic system, the patterns and dynamics present in that system are not always the same. Therefore, when the LSTM module misses a large leap, it makes a quick correction in time with new information, which in turn creates the curves in the Fig.~\ref{fig:results curves}.
 
To further evaluate the performance of our proposed model for short-term forecasting of cryptocurrencies, we also use the same sets of models for price forecasting of two other major cryptocurrencies, ETH and LTC. In this experiment, both the ETH and LTC data sets contain daily transaction data from April 15, 2016 to November 20, 2020. This includes the highest price, the lowest price, the bid price, the ask price and the volume, and all these data are from Quandl\footnote{https://www.quandl.com/}. Table~\ref{error2} shows the prediction errors under different DNN architectures. As this table shows, the results indicate that the LSTM outperforms the MLP in the time series task. The CNN-LSTM model uses the CNN for feature extraction before using the LSTM for prediction, which results in a smaller error than the LSTM. Finally, it is apparent from the table that the MRC-LSTM has the best performance. Our proposed MRC-LSTM model improves the feature extraction method by using a multi-scale residual convolutional layer for feature extraction, which gives the best prediction results in the experiments.

The empirical analysis shows that the proposed model also performs well in predicting the financial time series of the other two cryptocurrencies.

\section{Conclusion}
\label{sec:conclusion}
In this study, a hybrid method consisting of a multi-scale residual block and an LSTM network is proposed to predict the bitcoin price. Specifically, multi-scale residual block in this hybrid model is able to extract rich features at different time scales and also strengthen the representational ability of these features. Besides, the utilization of local residual learning leads to a reduction in computational complexity as well as improves the performance of the DNN. Different from some traditional studies, we introduce external influences such as macroeconomic variables and investor attention when performing price forecasting to combine various factors that may affect bitcoin prices. Furthermore, sufficient experimental results confirm that our proposed model has better prediction results than other single-structured models. The efficient feature extraction and integration capability of the proposed residual block is also demonstrated in the experimental comparison. In addition, we perform price prediction with the MRC-LSTM for two other major cryptocurrencies, ETH and LTC, further demonstrating the effectiveness of our model for multivariate time series forecasting of cryptocurrencies. In summary, compared with several state-of-art works, the proposed DNN model achieves better prediction performance, but it also has some limitations, since the Bitcoin market is particularly sensitive to some national policies, regulatory and market events, etc. More future work could be focused on comprehensive metrics which measure the investor's attention to more timely detection of bitcoin market volatility and thus more accurate price prediction.

\bibliographystyle{IEEEtran}
\bibliography{ref}

\end{document}